\def\beq{\begin{equation}}
\def\eeq{\end{equation}}
\def\bea{\begin{eqnarray}}
\def\eea{\end{eqnarray}}

\documentstyle[11pt]{article}

\def\double{\baselineskip 24pt \lineskip 10pt}
\input epsf

\begin{document}

\begin{titlepage} 
 
\vspace*{10pt} 
\begin{center} 
\LARGE 
{\bf Stellar footprints of a variable $G$\footnote{This essay received
an ``honorable mention'' in the 1999 Essay Competition of the Gravity
Research Foundation - Ed.}}\\
\vspace{0.3cm}
\normalsize 
\large{Diego F. Torres\footnote{Electronic address:
dtorres@venus.fisica.unlp.edu.ar}} \\ 
\normalsize 

\vspace{.2cm} 
{\it Departamento de F\'{\i}sica,  Universidad Nacional 
de La Plata, \\
C.C. 67, 1900 La Plata, Argentina}
\end{center} 
 
\vspace{.4 cm} 
\begin{abstract} 
\noindent  Theories with varying gravitational constant $G$
have  been  studied  since  long  time  ago. Among them, the most
promising  candidates  as  alternatives  of  the standard General
Relativity  are  known  as  scalar-tensor  theories. They provide
consistent descriptions of the  observed universe and arise as the
low energy  limit of  several pictures  of unified  interactions.
Therefore, an increasing interest on the astrophysical 
consequences of  such theories  has
been sparked over the last few  years. In this essay we comment
on two methodological 
approaches to study evolution of astrophysical objects
within a varying-$G$ theory, and the particular results we have obtained
for boson and white dwarf stars.
\end{abstract}

\end{titlepage} 

\newpage 
\double

\section{Introduction} 

\indent The idea of a varying gravitational constant $G$ has been in  the
physicist's minds since long time ago, when Dirac
proposed the large number hypothesis \cite{DIRAC}.  
It states that the  ubiquity of  some large
dimensionless  numbers  -${\cal O}(10^{40}$)-,  arising  from  the
combination  of  micro  and  macrophysical  parameters, is not a
coincidence but an output of an underlying time variation in
$e^2 G^{-1} m_p$, where $e$ is the unit charge and
$m_p$ is the proton mass.  
Well posed relativistic theories admitting time variation
in the fundamental constants of Nature had to wait for about 
twenty  years since  those  ideas,  until they were introduced,
in their present form,  
by Brans and Dicke \cite{BD}. 
The gravitational {\it ``constant''} then became a variable field. 

One of  the first  motivations to replace
General Relativity (GR) for a Brans-Dicke (BD)  or, more
generally,  for  a  scalar-tensor  (ST)  theory,  was  a  seeming
discrepancy   between   observations   and   the  weak  field  GR
predictions.  Time  went  by,  these
differences  vanished,  and 
the  main  motivation shifted to cosmology.

In ST  theories, the  gravitational action  has a  free parameter,
$\omega(\phi)$,  called  {\it   coupling  function}.  For particular 
values  of this parameter, these  theories
have cosmological  solutions which  are entirely  compatible with
all current gravitational  tests:  solar-system-based,
gravitational lensing, strong field pulsar tests,
and  nucleosynthesis \cite{WILL}.
Still, these theories may notably deviate
from GR, and might have played a crucial role in the early universe. 
The  scalar field may be a source of inflation and, 
moreover, some  ST theories are the low energy limit of unified pictures.
In general, we can say that, albeit severely  constrained,
a very slow time variation of $G$ cannot be discarded, especially  when
cosmological  time  intervals  are  considered. 
Others {\it constants} of Nature could well share this striking result
(see, for instance, the observational study on possible space-time 
variation of the fine structure constant \cite{ALPHA} 
and the consequent theoretical 
interest in a variable light speed \cite{LIGHT-SPEED}).

\section{Astrophysical modelling a variable $G$}

When $G$ is assumed to vary on cosmological intervals of time,
it is natural to expect that this will influence the evolution of all
non-transient astrophysical objects. 
A few years ago, the possibility of the existence of
{\it gravitational memory}  was introduced through the consideration of
what happens to black holes, during the 
evolution of the universe, if $G$ evolves with time \cite{GRAV-MEM}. 
One possibility is that the black hole evolves 
quasi-statically, in order to adjust its size to the changing $G$. If 
true, this means that there are no static black holes, even classically, 
during any period in which $G$ changes. Another possibility results when
the local value of $G$ within the black hole is 
preserved while the asymptotic value evolves with a 
cosmological rate. This would mean 
that the black hole remembers the strength of gravity at the 
moment of its formation. It
is immediate to extend this analysis to any object, like neutron
or other kind of stars. 

Then, the general problem we face 
is to have a complete solution, relativistic and space-time dependent,
for the metric, scalars, and matter fields, 
able to represent the evolution of a given object
through cosmic time. However, to
solve this problem is far from being an easy
task: in the usual applications, we assume fields that are 
either spatially-constant but time-varying (cosmology),
or spatially-varying but time-independent (astrophysics). 
Any real situation would require, however, a combination of both
and then, no global conservation law is in general avalaible.
Moreover, it worsens with the complexity of the internal 
structure of the object.

To consider these astrophysical scenarios,
we must adopt some simplifications. We shall explore two approaches:
either we consider simple stellar objects within a full
relativistic gravitation, or reduce the complexity in the
underlying theory (through Newtonian approximations), augmenting
that of the object under study. This latter
case might render direct observational consequences.

\begin{figure}[t]
\begin{center}
\leavevmode
\epsfxsize=10cm
\epsfysize=8.5cm
\epsffile{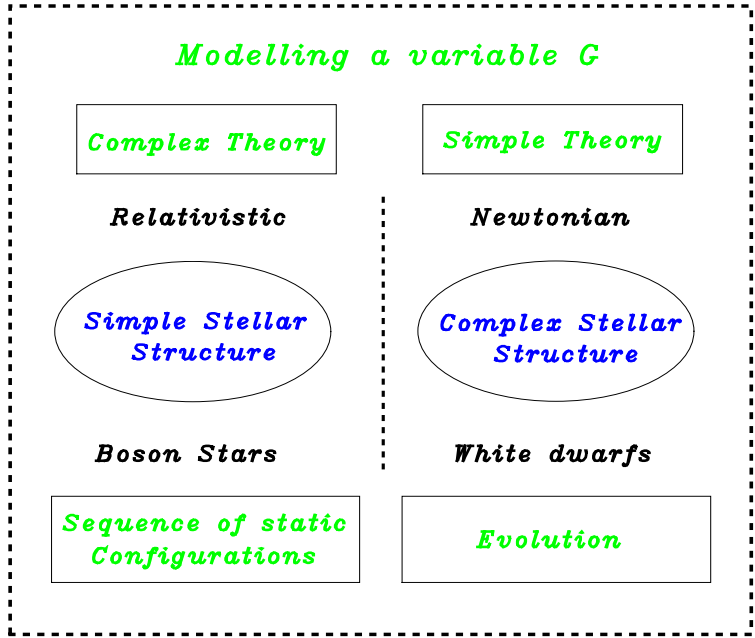}
\caption{Two different approaches to uncover the influence 
of a variable $G$ on astrophysical structures. In all real cases, $G$ varies 
in space because of the existence of the massive object,
and in time because the theory in which the object is modelled provides
a cosmological evolution. For usual situations, these two variations
are of the same order, amounting few percents of the present value of $G$.
The last box of both paths signals which kind of approximation
is involved. On the left, a relativistic treatment disregard an explicit
time dependence, and thus yields to sequences of static configurations.
On the right, the space dependence of $G$ is overwritten with a cosmological
time evolution, and as the treatment is Newtonian, we may compute
actual evolution of all matter fields.}
\end{center}
\end{figure}

In the left side of Fig. 1, we illustrate the first approach: we assume
a relativistic theory, say a scalar-tensor one, and
study simple theoretical constructs, with hardly
testeable predictions. 
We expect that the results so obtained may be at least
indicative of what happens with usual stars. If $G$ varies,
the value of the effective gravitational coupling far out
from the star must not necessarily be the Newton constant. 
On the contrary, it must take the 
value given by the evolution of a cosmological model
of the same gravitational theory (and at the same time)
in which the object is modelled.
This have the immediate consequence of changing the boundary conditions. 
It let us to approximately
study objects through different cosmic eras by starting from 
relativistic field equations and modifying the asymptotic value
of $G$; i.e. we are only capable to construct sequences of static  
configurations.\footnote{We cannot always state, however, 
that these sequences adequately represent
the real evolution of the object, as we can do below when considering
white dwarfs. In that case, because of appropriate approximations, 
the matter fields are already functions of time.}

We have studied this case for boson stars:
the analogue of a neutron star 
formed when a large collection of bosonic particles 
becomes gravitationally bound. Although 
such configurations were introduced in the 60's
\cite{KAUP-RB}, the current interest was prompted by the proof
that, provided the scalar field has a 
self-interaction, boson star masses could be of the same order of 
magnitude as, or even much greater than, the Chandrasekhar mass \cite{CSW}.
It seems possible for these stars to form through 
the collapse of a scalar field 
or gravitational cooling \cite{MLT}, 
though little is known yet. 
Since boson stars are easier to compute 
than black holes, because of
the absence of singularities and horizons, we were able to numerically
solve their equilibrium structure in several scalar-tensor theories,
for different cosmic times.
We have explicitly shown that these configurations 
are sensitive to small variations in the boundary condition for $G$,
these changes amounting for a few 
percents of their own masses. Assuming the past value of the coupling 
constant $G$ to be greater than the present one,
we found that the mass and particle number of
static configurations, at fixed central density, increase from 
earlier to later times. Although stable stars may exist at all
times, as time goes by, 
models with a given central density moves towards the stable branch.
Also, the radius-mass relationship 
is appreciably modified. Our latest results, assuming that the system evolves
conserving the number of particles, show that the mass
and the central density must decrease in time (considering a constant
mass unit). The reduction in central density seems reasonable when
we recall the force balance in polytropic stars: 
since gravity is reducing in
strengh, the equilibrium configurations can become more diffuse and hence
drop in central density.
Detailed discussion of these results 
may be found in Refs. \cite{BOSON}.
All in all, despite we are not able to answer which phenomenon
(memory or quasi-static evolution) actually happen, 
we confirm that these equilibrium spheres 
are sensitive to a very slow cosmological variation of $G$. 

To judge whether a quasi-static evolution is 
a feasible scenario, we can compare the free-fall time 
of the stellar structure with the rate of $G$ variation. The free-fall time
(also known as the hydrostatic time-scale, $\tau_{ff} \sim 1/\sqrt{G \rho}$,
with $\rho$ the density of the object) gives the typical scale 
in which a dynamical stable star reacts
to a slight perturbation of hydrostatic equilibrium. In the case of usual
stars, this scale is extremely short: 
it is of order of minutes for a star like
the Sun and of order of seconds for white dwarfs (WDs). This is negligible
when compared with the rate of variation of $G$, which occurs over a
scale similar to that of the age of the universe.
Then, in most phases of the star evolution, one may safely consider them
in hydrostatic equilibrium; and so would be even if $G$ is 
a variable funtion.

WDs, in particular, do not generate gravitational fields too large
as to necessarily consider them as relativistic objects. Retaining terms
up to the post-Newtonian approximation, the first correction to the Newtonian
equilibrium is proportional to $P/\rho c^2$, largely less than 1
for typical star pressures and densities. Then, we may adopt the
methodological procedure of the
right side of Fig. 1: take a simple theory and a complex object, 
and try to isolate observable effects. In this case,
we are able to see the evolution of the {\it same} object through cosmic
time, because under the Newtonian hydrostatic equilibrium approximation,
evolution is just a sequence of static configurations.

\mbox{}From the above considerations, 
it is natural to expect that the direct introduction of a time
varying $G$ into the equations that represent
the equilibrium structure of WDs will be a
safe procedure. The star will be able to see that variation, reacting to it
immediately: it can not remember the value of $G$ at its formation
but it is forced to evolve changing with it. A question still
remains: if the rate of change in $G$ is slow enough as to agree with other
gravity tests, are these changes appreciable? 

Indeed, if $G$ varies, WDs evolution could be sensitive enough as to provide 
a good independent method for measuring its change. 
There are two immediate reasons for this: firstly, they have 
lived during most of the life of the universe, and have
time to integrate extremely small values of the rate 
of change of $G$. In addition,
at their latest stages of evolution, their luminosity arise from a delicate
balance between gravitational and thermal energies, and any change in $G$
might strongly affect that equilibrium. This may produce a different
luminosity function.\footnote{The 
luminosity function (WDLF) is defined as the number of WDs with 
a given luminosity per cubic
parsec and per unit luminosity.}

We have run a detailed stellar evolution
numerical code, with state-of-the-art
physical inputs corresponding
to WDs at each time of their evolution, and with a careful account of
a variable $G$. Making no {\it a priori} assumptions
on the thermal behavior of the interior of the WD,
we  have computed  the  evolution  of  C/O (hydrogen envelopes) 
models with masses ranging from $M=0.4M_\odot$
to $M= 1.0M_\odot$,  at
intervals  of  0.1$M_\odot$, and with  zero  metallicity. We
assumed the mass of the hydrogen and helium layers to be  $M_{\rm
H}/M = 10^{-5}$ and  $M_{\rm He}/M = 0.01$, and 
a decreasing value of $G$. We took $G \propto
t^{-2/4+3\omega}$, with $\omega$ a constant, which is the cosmological
solution of the BD theory in the matter era. Details of the simulations,
the code, and the input physics may be found in Ref. \cite{WD}
and references cited therein.

The first striking result is that, because of the presence of 
a variable $G$ in the energetic balance equations,
the cooling of WDs is strongly
accelerated, particularly at low luminosities. Owing to the  fact
that more massive WDs have  smaller radii, this effect turns  out
to be far more dramatic in massive WDs.
\begin{figure}[t]
\begin{center}
\leavevmode
\epsfxsize=8cm \epsfbox{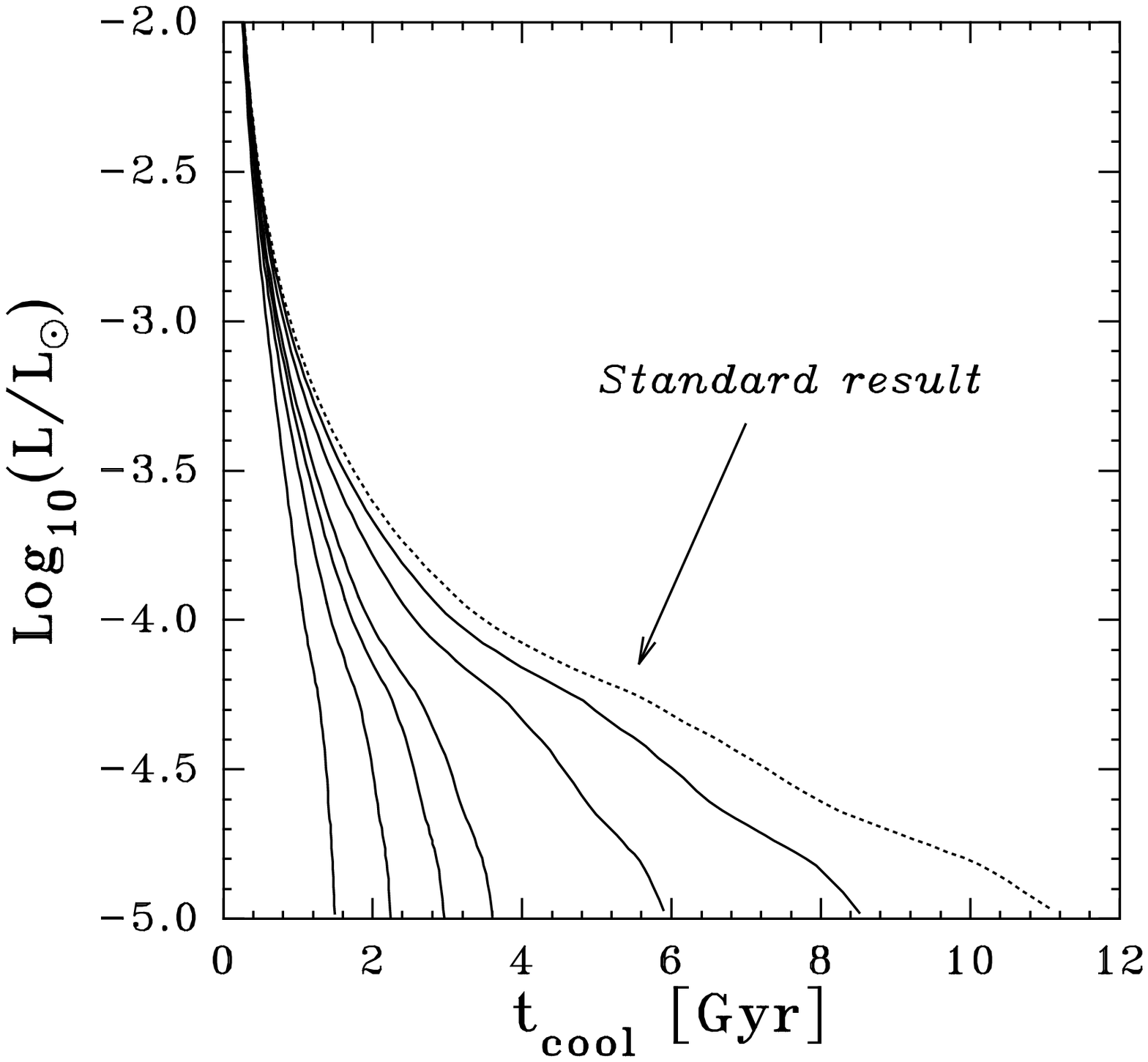} \hspace{0.1cm} 
\epsfxsize=8cm \epsfbox{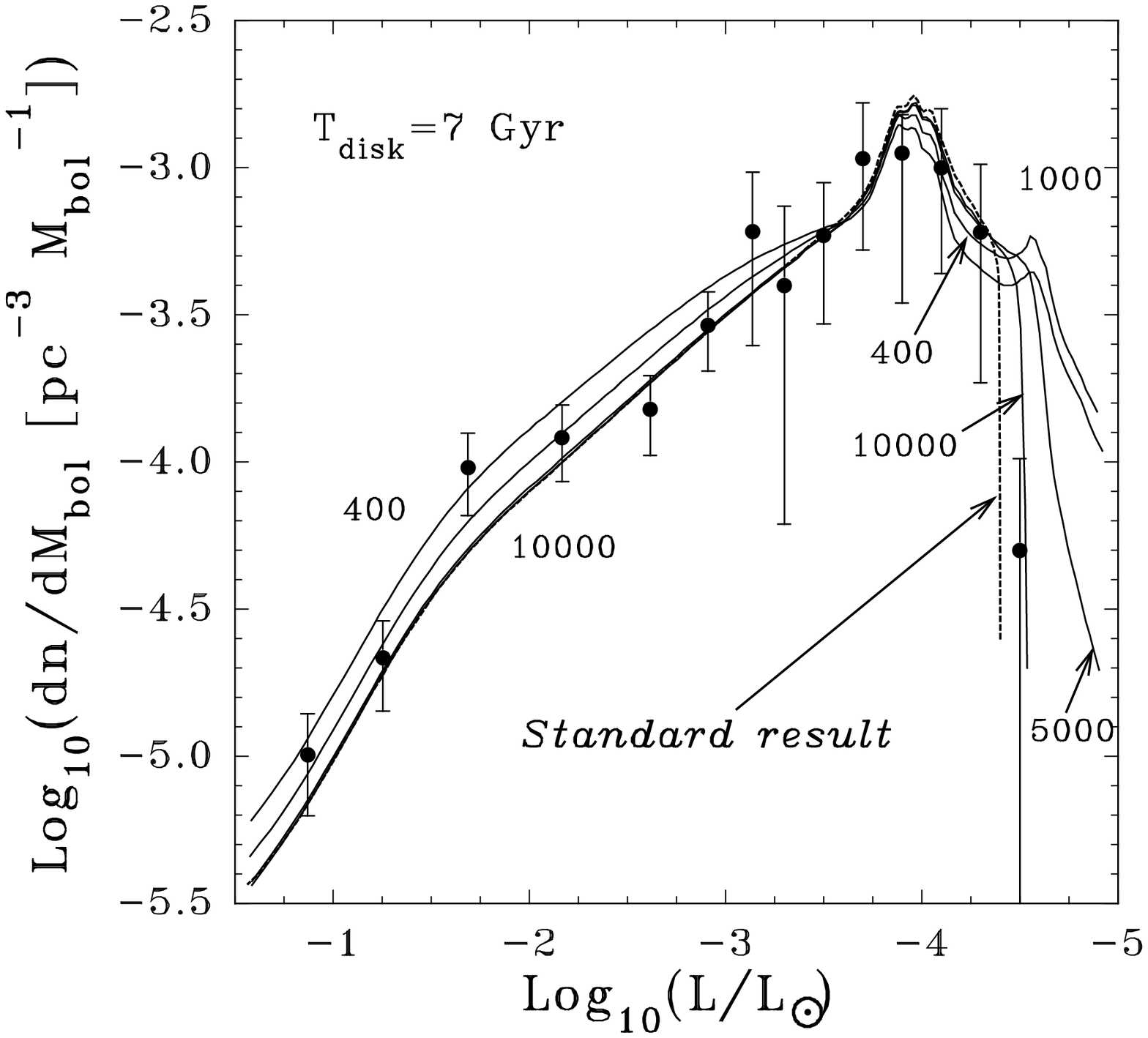}\\
\end{center}
\caption{The figure on the left shows the acceleration of the cooling of
a 0.6$M_\odot$ WD assuming an age for the universe of 12.5
Gyr for different values of the $\omega$ parameter, from left to
right, going from 400 to 5000 ($G \propto t^{-2/4+3\omega}$). The figure
on the right shows the observed points of the luminosity function
together with the theoretically computed ones. The dashed line corresponds
to the $G$-constant solution 
and the solid ones to the $G$-varying
case, with $\omega$ ranging from 400 to $10^3$. The galactic disk age was
assumed to be 7 Gyr, possibly the lower plausible limit. 
Bolometric magnitudes are $M_{bol}=-2.5 \log (L/L_\odot) + 4.75 $.
Observed points are taken from
Leggett et al.,  Ap. J. 497, 294 (1998). The theoretical computation
of the luminosity function follows Iben and Laughlin,  
Ap. J. 341, 312 (1989).}
\end{figure}
In Fig. 2 (left) we  show the  relationship between the
logarithm of the luminosity of the  WD and its age. 
The effect of $\dot{G} \neq 0$ is noticeable: while for very
high values of $\omega$ the function is rather similar to that of  the
standard case, the time spent  by a 0.6 $M_\odot$ WD  in reaching
$\log{L/L_{\odot}}= -5$ fall to  a fifth for $\omega=$1000. Not
surprisingly, for higher  stellar masses,  these differences
become larger. In Fig. 2 (right) we show the theoretical WDLF 
along with the observed points. Amazingly, we find
that even  considering a  value of  $\omega$ as  large as 10$^3$,
differences are found between the 
$\dot G\neq 0$  and $\dot{G}\equiv 0$ behaviors. 
Values of $\omega  < 5000$ should be discarded,  as we
see no agreement between observed and computed WDLFs 
(recall the solar system limits $|\omega| > 500$). 
That would  imply  a  current   value  of  $|\dot{G}/  G|$   of  order
$10^{-14}$, which  is between  1 and  3 orders  of magnitude more
restricitive than previous  bounds.\footnote{This limit is valid for our
cosmic time because the WDLF is constructed with stars in our neighbourhood,
all of them seeing our current value of $G$.}
It  is  important to stress that the lowest
luminosity point  in the  observed distribution  is the
least precisely determined,
and it is still quite possible that its
position may  suffer variations  in the  near future.

On the basis  of these results,  we conclude that  the evolution of
WDs (even in the context of the assumed approximations) is
a very powerful tool to probe the variation in the value  of $G$ with a  
greater degree of sensitivity than that provided by other
experiments.  When a comprehensive knowledge of the WDLF
be avalaible,  a comparison with  these results
may help to decide whether or not a $\dot G \equiv
0$ theory is a better description of gravitation.

\section{Final comments}

We have intended to analyze two approximate schemes 
to an unique problem: the
determination of the influence of a varying-$G$ cosmology 
upon astrophysical structures. 
This would mean to exactly solve relativistic field
equations, and simultaneously provide the behavior of the matter 
and the metric (together with the scalar related with $G$). 
The real situation
would imply space and cosmic time dependences, and it is hardly tractable.
However, as soon as the theory predicts a cosmological variation of $G$, 
the real evolution of the objects will also require a running $G$-value
(unless memory effects are operative). By following the two 
methodological lines depicted in Fig. 1, 
we have studied these phenomena for boson and WD
stars. 

WD cooling was recently studied in Ref. \cite{BERRO}, with the same 
energetic balance input and similar qualitative
results. However, the crucial assumption of Ref. \cite{BERRO}
was an isothermal interior, something that we explicitly 
found as not valid. This absence of isothermicity 
produces an even stronger acceleration of the 
cooling and more notorious observational effects in the luminosity
function. Time variation of $G$ was also analyzed as the possible cause
of the discrepancy between the cosmic expansion age and the apparent
globular cluster age \cite{RAFFELT}. Observational signatures
of boson stars are currently being investigated \cite{BS-S}, and the 
possibility of using these as discriminators of different
theories of gravity is under analysis, especially in microlensing
phenomena \cite{AWW}.

In static situations, the most dramatic effect recently discovered
is spontaneous scalarization \cite{PRL}. It happens in neutron
stars models in general scalar-tensor theories, and provides nonperturbative
effects which induce large deviations from GR behavior. The gravitational
equilibrium configuration  of a neutron star -at a given cosmic time-
suffers spectacular changes, increasing their maximum masses. This
effect has impact on neutron star binary coalescence and pulsar tests.

A comprehensive knowledge of all the influence that a varying-$G$
cosmology would have on astrophysics seems to be far,
especially when one considers the large class of
astrophysical objects with long lifetimes, and not just isolated
stars. It is certain, however, that studying this problem
provides us with the possibility to gain a deep insight of  
fundamental physics, and that it is worth exploring.

\section*{Acknowledgments}

It is a pleasure to thank L. G. Althaus,
O. G. Benvenuto, A. R. Liddle, F. E. Schunck, and A. W. Whinnett for the different
collaborations which made possible the results commented here,
and to L. Anchordoqui and G. E. Romero
for a critical reading of this
essay. The research on which this essay is based was supported by
CONICET (Argentina), the British Council (UK), while the author
was a Chevening Scholar, and Fundaci\'on Antorchas (Argentina) under grant
A-13622/1-98.

\end{document}